\documentclass[english,aps,preprint]{revtex4}
\usepackage[T1]{fontenc}
\usepackage[latin9]{inputenc}
\usepackage{amsmath}
\usepackage{graphicx}
\usepackage{esint}

\makeatletter
\@ifundefined{textcolor}{}
{%
 \definecolor{BLACK}{gray}{0}
 \definecolor{WHITE}{gray}{1}
 \definecolor{RED}{rgb}{1,0,0}
 \definecolor{GREEN}{rgb}{0,1,0}
 \definecolor{BLUE}{rgb}{0,0,1}
 \definecolor{CYAN}{cmyk}{1,0,0,0}
 \definecolor{MAGENTA}{cmyk}{0,1,0,0}
 \definecolor{YELLOW}{cmyk}{0,0,1,0}
 }

\makeatother

\usepackage{babel}

\begin{document}

\title{Isochronic Pendulum}

\author{Sumit kumar}
\begin{abstract}
\noindent The classic simple pendulum is a device which works as a
simple harmonic oscillator (S.H.M.) only approximately. The time period
remains fixed as long as the amplitude is kept sufficiently small.
This limitation makes it unsatisfactory choice for practical time
keeping purposes. The question addressed in this paper is regarding
the modifications that can be made to the pendulum so that its time-period
is independent of amplitude. Though the idea is not new, it is shown
using rigorous mathematics that if the length of the suspension is
suitably controlled so as to have a cycloidal path for the bob, this
can be achieved.
\end{abstract}
\maketitle

\section{introduction}

The classic pendulum problem is stated in the following way: A point
mass is suspended with a massless inextensible string under gravity
with acceleration g and released from a point different from the mean
point with the string kept tight. The equation of motion is a second
order differential equation. Hence time period is dependent on the
amplitude. Under the assumption that the displacements from the mean
position are small, the time period becomes independent of the amplitude
resembling a simple harmonic system.

\section{THE CLASSIC SIMPLE PENDULUM}

The standard pendulum problem is described by the Lagrangian:\begin{equation}
L=\frac{1}{2}\mu l^{2}\dot{\theta^{2}}+\mu gl\cos\theta\end{equation}

\noindent The equation of motion becomes:\begin{equation}
\ddot{\theta}=-\dfrac{g}{l}sin\theta\end{equation}

This equation gives the solutions for $\theta$ . The time period
for the motion is in general \emph{not} a constant. However, if we
restrict $\theta$ to be small, we have sin$\theta\sim\theta$, so
the approximate equation of motion becomes:\begin{equation}
\ddot{\theta}=-\dfrac{g}{l}\theta\end{equation}

This equation represents the simple harmonic oscillator with a time
period:\begin{equation}
T=2\pi\sqrt{\dfrac{l}{g}}\end{equation}

which is clearly independent of $\theta$ for a fixed length $l$
. This is a well known result.

Now the question is- why do we need to modify the design ?

It turns out that if we use the non-linear solution for $\theta$,
the time period increases with increase in $\theta$. The total length
of the pendulum is fixed. However we might be able to make suitable
modifications to the system so as to make time period independent
of $\theta$. There should be some way to manipulate this length with
$\theta$ so that the time period remains fixed. In general this will
turn out to be a function between $\theta$ and a parameter which
will behave as {}``effective'' length at a particular point. So
we need some external mechanism to achieve the required effect.

\section{MODIFICATION TO THE CLASSIC SIMPLE PENDULUM}

According to the original simple pendulum, the trajectory of the particle
is a circular arc. It is \emph{this curve }which makes the time period
a function of $\theta$ . If we want to rectify this, we should assume
that the particle follows a generic curve. So our task is to find
the equation of this curve under the constraints that the particle
be in S.H.M. This condition will guarantee the time period to be a
fixed quantity.

The equation of a point particle under S.H.M is given by\begin{equation}
\ddot{\xi}=-k\xi\end{equation}

where $\xi$ is the displacement along the generic curve, and k is
some constant to be fixed according to the system.

From the Pythagoras theorem, we have\begin{equation}
(d\xi)^{2}=(dx)^{2}+(dy)^{2}\end{equation}

\begin{equation}
\xi=\int\sqrt{(dx)^{2}+(dy)^{2}}\end{equation}

Since the body is under S.H.M, it is acted upon by a potential of
the form:\begin{equation}
P.E.=\frac{1}{2}k\xi^{2}\end{equation}

For a particle under gravity the potential is $\mu gy$ where $y$
is the vertical component of position from the reference.

So we have\begin{equation}
\mu gy=\frac{1}{2}k\xi^{2}\end{equation}

Using eqn. (7) we get\begin{equation}
\mu gy=\frac{1}{2}k\left[\int\sqrt{1+\left(\dfrac{dx}{dy}\right)^{2}}dy\right]^{2}\end{equation}

Differentiating w.r.t $y$ and squaring on both sides, we get

\begin{equation}
\Rightarrow\frac{dy}{dx}=\sqrt{\frac{2ky}{\mu g-2ky}}\end{equation}

To simplify this equation, we use a following substitution:

Let\begin{equation}
\dfrac{dy}{dx}=\tan\varphi\end{equation}

Then using eqn. (11) we get\begin{equation}
\sin\varphi=\sqrt{\frac{2ky}{\mu g}}\end{equation}

Squaring and using basic trigonometry we get\begin{equation}
y=r\left(1-\cos2\varphi\right)\end{equation}

Where $r=\dfrac{\mu g}{4k}$.

Similarly we also have\begin{equation}
\frac{d\varphi}{dx}=\frac{dy/dx}{dy/d\varphi}\end{equation}

which gives us\begin{equation}
x=r\left(2\varphi+\sin2\varphi+2c\right)\end{equation}

Here $c$ is a constant of integration. After a simple change of variables
and fixing the integration constants we can get the following equations:\begin{equation}
x=r\left(\sin\theta-\theta\right),\, y=r\left(1+\cos\theta\right)\end{equation}

These are the standard equations of $cycloid$. Hence, if a particle
\emph{tied to a fixed point} executes a simple harmonic motion under
the action of gravity, it must follow a trajectory of a cycloid.

The mechanism to achieve this effect will be discussed later.

Let's calculate the time taken by a point particle following this
trajectory from any initial height $y_{0}.$

Time taken by a particle is given by\begin{equation}
t=\int\frac{d\xi}{v}\end{equation}

$d\xi$ is the differential path traveled and $v$ is the velocity
of the particle which is a function of $y$.

From the conservation of energy we have\begin{equation}
\frac{1}{2}\mu v^{2}=\mu g\left(y_{0}-y\right)\end{equation}

where $y_{0}$ is the initial height. It is assumed that $y=0$ is
the reference.

So, $v=\sqrt{2g\left(y_{0}-y\right)}$

Using this in eqn. (18) we get\begin{equation}
t=\int\limits _{y_{0}}^{0}\sqrt{\frac{(dx)^{2}+(dy)^{2}}{2g\left(y_{0}-y\right)}}\end{equation}

Now, using eqn. (17) and after simplification we get\begin{equation}
t=\sqrt{\frac{r}{g}}\int\limits _{\theta_{0}}^{\pi}\left(\frac{1-\cos\theta}{\cos\theta_{0}-\cos\theta}\right)d\theta\end{equation}

Here $\theta_{0}$ and $\pi$ are the angles corresponding to $y_{0}$
and $0$.

Solving this integral we get the $t=\pi\sqrt{r/g}$.

Total time period will be four times $t$. Hence\begin{equation}
T=4\pi\sqrt{\frac{r}{g}}\end{equation}

This shows that time period is independent of the initial height and
hence independent of the amplitude.

Here we see that the time-period depends upon $r$ , which looks like
an {}``effective length'' of the {}``simple'' pendulum. So we
see that the Length of the {}``simple'' pendulum needs to be changed
with change in amplitude. This is an expected result.

Let's discuss the mechanism to make a pendulum bob follow the trajectory
of a cycloid. The answer to our problem lies in the elementary theory
of curves.

\section{DESIGNING THE PENDULUM}

From the elementary theory of curves the parametric equation of an
evolute is\begin{equation}
\alpha=x-\frac{\left(1+y'^{2}\right)y^{'}}{y''},\,\beta=y+\frac{1+y'^{2}}{y''}\end{equation}

Here $x$ is the parameter and $y$ is a function of $x$ .

Substituting eqn. (17) in eqn. (23) we get the following equations:\begin{equation}
\alpha=-r(\sin\theta+\theta),\,\beta=r(3-\cos\theta)\end{equation}

Making a simple substitution $\theta=\pi+\theta'$ we get\begin{equation}
x'=r(\sin\theta'-\theta'),\, y'=r(1+\cos\theta')\end{equation}

\noindent where $x'=\alpha+r\pi$ and $y'=\beta-2r$.

\includegraphics[scale=1.25]{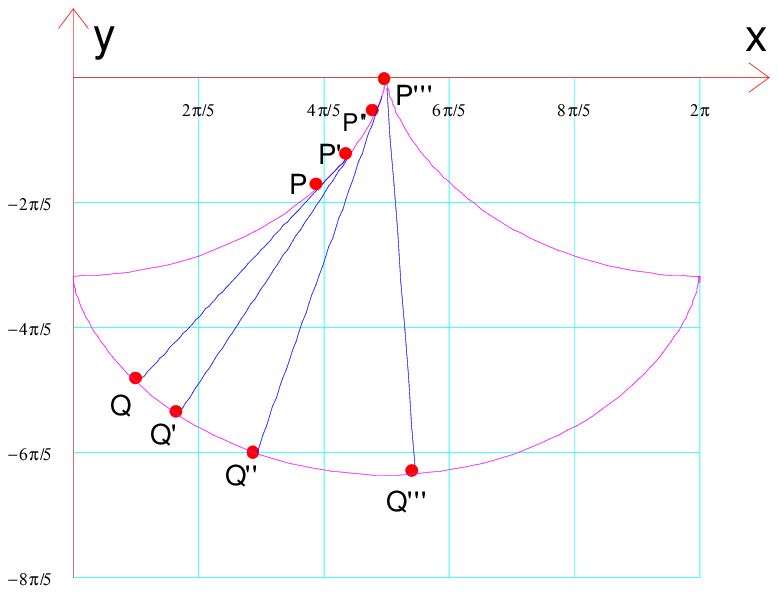}

This is a remarkable result !

It shows that evolute of a cycloid is a $cycloid$ itself ( apart
from the overall coordinate shifts). This can be done using the following
simple construction.

This analysis shows that if the tangents are drawn to the cycloid
at various points $Q$, $Q'$, $Q''$, the locus of the end- points
$P$, $P'$, $P''$ is also a cycloid. The length of the tangents
decreases and it turns out that this variation exactly compensates
for the dependence of time period with amplitude. Hence, using this
simple modification, we can build a pendulum whose time-period is
constant even when the amplitude of oscillations are large.

It decreases in a specific way due to the curve of cycloid. This gives
us some ideas. If we somehow manage to make the bob follow the locus
of the end-points of these tangents, our problem will be solved. This
turns out to be a simple task.

We can make a rigid mechanical barrier in the shape of a cycloid with
frictionless grooves for the string to move freely along the cycloidal
barrier. This should be done on both sides (see the figure below)
to allow a periodic motion of the bob. The physical parameters for
the construction of the mechanical barrier can be determined for a
fixed value of time period. The choice of the initial values of $\theta'$
will affect the motion of the bob and also the physical dimensions
of the barriers. The calculated values of $r$ should be appropriately
matched with the initial value of $\theta'$ so that the motion of
the bob is not too rapid and wide. Also, one should be careful while
choosing the initial length of the pendulum so that the bob does not
hit the mechanical barriers.

\noindent \begin{center}
\includegraphics[scale=0.5]{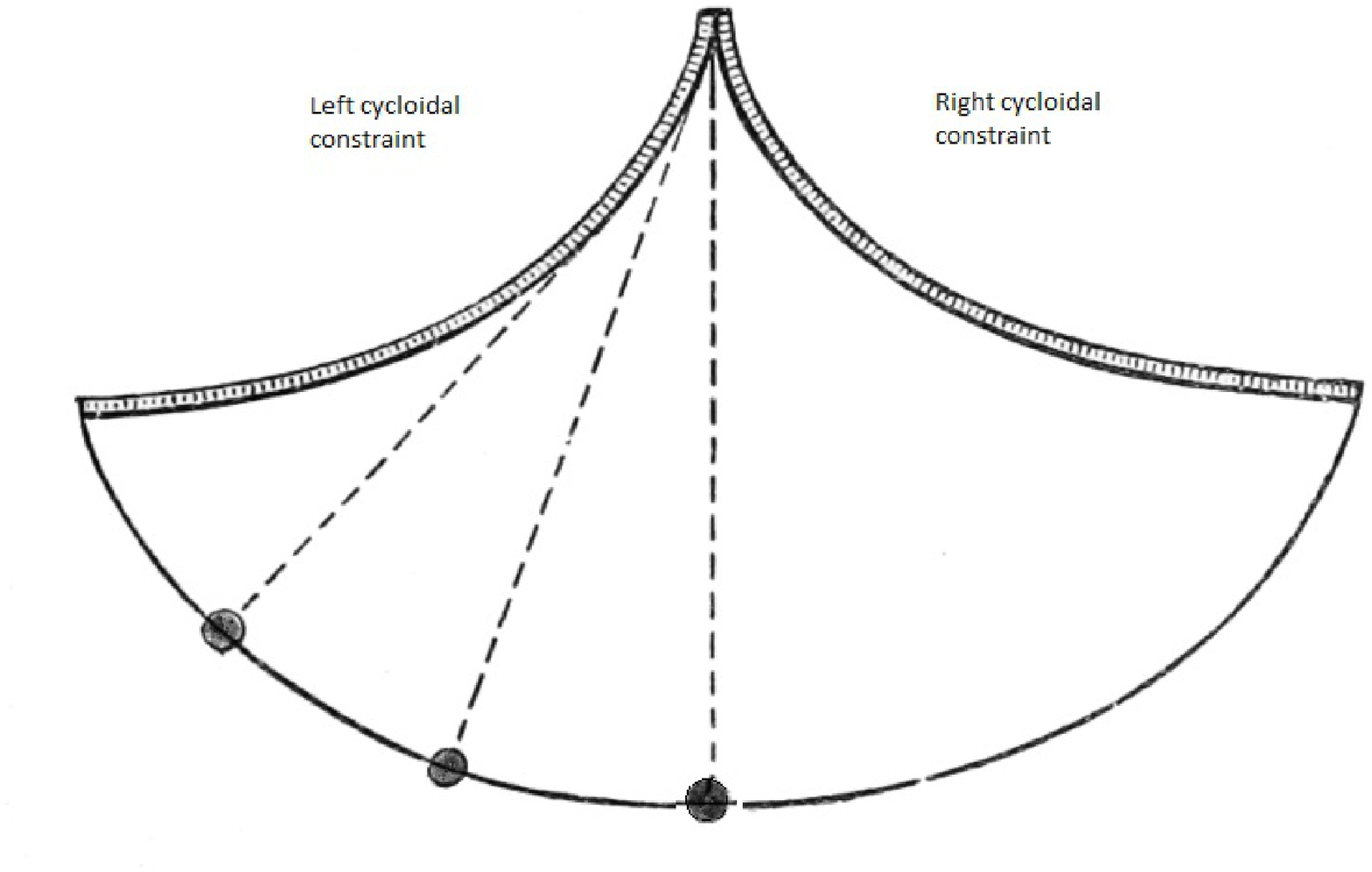} 
\par\end{center}

\noindent \begin{center}
The modified simple pendulum 
\par\end{center}

\section{CONCLUSION}

A simple pendulum is just an approximation to S.H.M. So in general
it does not have a time-period which is independent of its amplitude.In
this paper it is shown that this problem can be solved in a simple
way by making the pendulum bob follow a cycloidal trajectory using
cycloidal constraints and a practical design to achieve the constraint
is also suggested. Hence the new design is free from any approximation
and serves well for the purpose of fixed time-periods.

\end{document}